\documentclass[manuscript,screen,sigconf,nonacm]{acmart}

\usepackage{todonotes}
\usepackage{float}
\usepackage{algorithm}
\usepackage[noend]{algpseudocode}
\usepackage{xcolor}
\usepackage{tcolorbox}
\usepackage{pgf}
\usepackage{mdframed}
\usepackage{expl3}
\usepackage{environ}
\usepackage{etoolbox}
\usepackage{tikz}
\usepackage{pgfplots}
\usepackage{xcolor}

\usepackage[inline]{enumitem}
\ExplSyntaxOn
\seq_new:N \g_appendices_seq
\NewEnviron{Appendix}{\seq_gput_right:No \g_appendices_seq \BODY}
\newcommand\AddAppendices{
  \onecolumn
  \appendix
  \seq_map_inline:Nn \g_appendices_seq {##1}
}
\ExplSyntaxOff

\AtEndDocument{\AddAppendices}
\begin{document}


\title{Exploring the Potential of Conversational Test Suite Based Program Repair on SWE-bench}

\author{Anton Cheshkov, Pavel Zadorozhny, Rodion Levichev, Evgeny Maslov, Ronaldo Franco Jaldin}


\begin{abstract}
Automatic program repair at project level  may open yet to be seen opportunities in various fields of human activity. Since the SWE-Bench challenge was presented, we have seen numerous of solutions. 

Patch generation is a part of program repair, and test suite-based conversational patch generation has proven its effectiveness. However, the potential of conversational patch generation has not yet specifically estimated on SWE-Bench. This study reports experimental results aimed at evaluating the individual effectiveness of conversational patch generation on problems from SWE-Bench.

The experiments show that a simple conversational pipeline based on LLaMA 3.1 70B can generate valid patches in 47\% of cases, which is comparable to the state-of-the-art in program repair on SWE-Bench.
 
\end{abstract}

\maketitle

\section{Introduction}

The grows of AI capabilities opens new horizons for automation of software engineering tasks. As an illustration, the SWE-Bench benchmark \cite{jimenez2024swebenchlanguagemodelsresolve} was recently published to challenge and evaluate AI ability to accomplish Automatic Issue Resolving (AIR) task.

The majority of AIR approaches explicitly or implicitly leverage program repair techniques such as \textit{fault localization}, \textit{patch generation}, and \textit{patch validation}. LLM-based Conversational Patch Generation (CPG), that runs tests and uses failing information in dialog, has shown its advantage \cite{xia2023conversationgoingfixing162, zhang2024criticalreviewlargelanguage} over repetitive patch generation. 


To date, the leading AIR approaches\footnote{ \url{https://swebench.com}} may resolve at least 43\% of tasks in SWE-Bench Lite.  However, what is not yet known is the potential of CPG as a part of AIR system design. 
If one had \textit{oracle fault localization}\footnote{The localization that can be used to repair the defect} and a unit test, what would the effectiveness of the repair be? Knowing the potential of each individual component within the system allows planning the improvement of the entire system.


The aim of this study is to estimate the potential of LLM-based conversational patch generation. We conduct an experiment using problems from SWE-Bench benchmark. In our experiment the function-level fault localization is known and the failure information of running test suite is used in the repair dialog with LLM.

We want to answer the following research questions:

\begin{enumerate}
    \item What is the effectiveness of conversational patch generation with a test failure information feedback on SWE-Bench
    \item Is conversational patch generation with a test failure information feedback more effective than regular repetitive LLM-base patch generation at the same budget of requests to the LLM
\end{enumerate}




\section{Methodology}

In order to assess the potential of LLM-based conversational program repair in the context, present study adopts Automatic Issue Resolving (AIR) problems from SWE-Bench dataset. For each SWE-Bench record, the fault localization\footnote{Can be extracted from the gold patch} and failing tests are used to start conversational repair pipeline. The repair process is considered successful when all previously failing tests pass.

In the original AIR task, the fault localization and reproducible example are not given. The proposed experiment allows to estimate the potential of conversational program repair, but can not be used for comparison with state-of-the-art approaches in AIR task.

The first step in our experiment is to prepare dataset of programs, defect descriptions, and failing test suites. After, conversational program repair is run for each program. Finally, the effectiveness metrics are calculated.

\subsubsection*{Task Definition}

Let $P$ is a GitHub project, that have an issue $i$ described in natural language, and there is a set of failing unit-tests $T$ because of the issue $I$, and a function $s$ from $P$. The task is to find a $\hat{s}$ the new version of $s$, that if replace $s$ with $\hat{s}$ it ensures that all tests in $T$ pass. 

The defined task is similar to test suite based repair except the precise fault localization $s$ is provided, along with the issue description $i$.


\subsubsection*{Data}\label{sec:data}

The SWEBench is a collection of GitHub issues and corresponding patches from 11 well-known open-source python projects. The Lite subset of SWEBench contains 300 issues where the gold repair patch affects only a single project's file\footnote{not including changed in unit tests}. 

Initially, we pick out 192 problems from SWEBench Lite where the gold patch is localized in a single function. We exclude all Django problems\footnote{Because its unit test framework does not allow to run specific set of tests, and outputs test failure information in a non-structured way}. Additionally, we manually validated each problem environment to be sure that all unit tests mentioned in the benchmark can be unambiguously found in test framework output logs. Finally, it results into 92 problems\footnote{For some problems we did not managed to setup environment.}; the full list can be found it the Appendix [].

From each of the chosen problems we extract the following information: 
\begin{enumerate*}
    \item issue description $i$;
    \item faulty function $s$;
    \item public test site $T$;
    \item hidden test site $T^*$.
\end{enumerate*}

The faulty function is obtained from the oracle patch. The test suites $T$ and $T^*$ corresponds to FAIL\_TO\_PASS\footnote{All tests have failed status} and PASS\_TO\_PASS\footnote{All tests have passed status} fields in SWEBench.

\subsubsection*{Program Repair Conversation}
A conversation repair algorithm is a question-answer dialog between a developer and LLM. Each message from the developer to LLM requests to generate a repaired version of the program.

The developer starts conversation by a message comprised of the faulty program $s$ and issue description $i$ combined in the message template $MSG_A$. The LLM response is parsed to make a patch, apply it, and run the test suite $T$. Any issue that does not allow successfully execute the test suite $T$ is considered as a syntax error and the constant error $MSG_B$ message is sent back to LLM. In case, test suite is successfully run, but not all tests from $T$ are passed the message $MSG_C$ with appended test failure information is sent to LLM.

The conversation continues till either all tests from $T$ are passed or the number of messages to LLM reach number of attempts. The formal description of conversation written in Algorithm \ref{alg:conversation}.

Study runs two variants of experiment. The first runs 6 consecutive independent conversation rounds with 5 attempts limit each. The second runs 30 independent conversations within 1 attempt. The last experiment is a equivalent to repetitive LLM-based repair without failure tests feedback. Both experiments calls LLM the same amount of times.
\subsubsection*{Metric Choice}

The repair conversation is considered to be successful if the generated patch passes public test suite $T$. We measure the effectiveness as percent of successful conversations among those SWEBench problems we chose (see Section \ref{sec:data}).

In the SWE-Bench both $T$ and $T^*$ test suites are used for patch validation. Therefore, as soon as the plausible patch is found, the hidden suite $T^*$ is also run. This information allows to calculate the  metric of successfully resolved issues as it was proposed in original SWE-Bench \cite{jimenez2024swebenchlanguagemodelsresolve}.

Our metrics cannot be strictly used to compare to any AIR approaches because we are addressing different tasks and utilizing different inputs.

\subsubsection*{LLMs choice}

The study is not aimed to evaluate and compare LLMs between each other.
The choice is constrained by the time, available hardware, and experiment budget. Eventually, one open-sources and one proprietary models have been chosen.  

The first model is llama3.1 70B Instruct quantized to 4 bits. This model was deployed on a the server with 48GB of GPU and serves over HTTP API provided be ollama server []. The second choice is close-sourced gpt-4o-mini that runs over HTTP API.

\begin{algorithm}
\caption{Conversational Program Repair}\label{alg:conversation}
\textbf{Input:}
 {\begin{minipage}[t]{8cm}%
     \begin{tabular}{ll}
       $s$ & Code snippet to repair \\
       $i$ & Problem description \\
       $T$ & Set of unit tests \\
       $M$ & LLM program interface \\
       $attempts$ & Number of conversation iterations \\
       $MSG$ & Set of message templates of 3 elements
    \end{tabular}
   \end{minipage}%
  }
\\
\begin{algorithmic}
\State $H \gets [MSG_A(s, i)]$ \Comment{init chat history}
\While{$attempts > 1$}
\State $out  \gets M.chat(H)$
\State $res \gets try\_apply\_patch\_and\_run\_tests(out, T)$
\State \textbf{case} $res$ \textbf{of}:
\State \hspace{\algorithmicindent}  SyntaxError: $H \gets append\:MSG_B\:H$
\State \hspace{\algorithmicindent}  SemanticError $err$: $H \gets append\:MSG_C(err)\:H$
\State \hspace{\algorithmicindent} otherwise: break \Comment{all tests are passed}
\State $attempts \gets attempts - 1$
\EndWhile
\end{algorithmic}
\end{algorithm}

\begin{figure}[htbp]
    \centering
    \resizebox{\columnwidth}{!}{
        \includegraphics{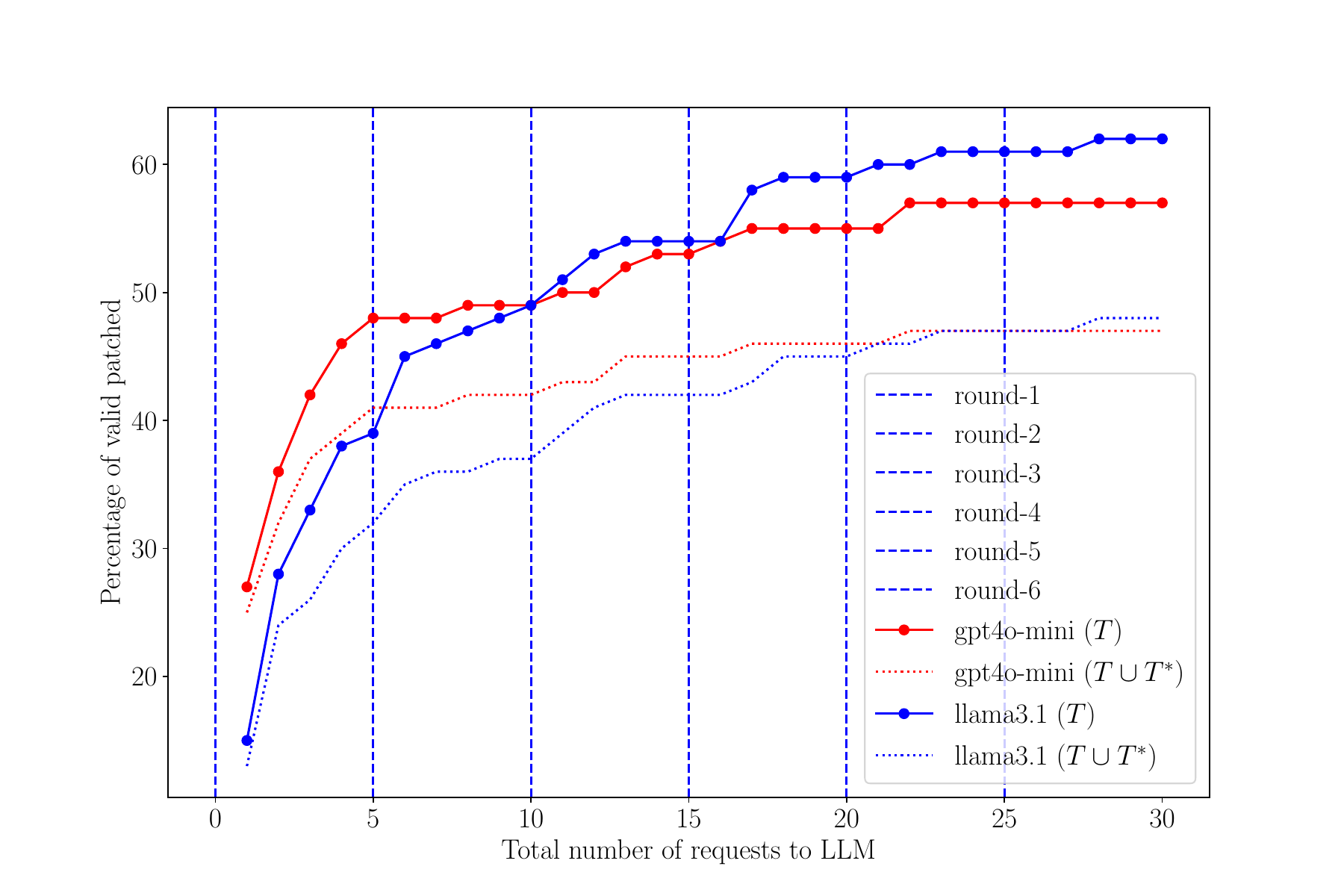}
    }
    \caption{Accumulated percent of valid patches generated during 6 consecutive independent conversations, 5 LLM requests each. Two LLMs: llama3.1 70B Instruct and gpt4o-mini. Two different patch validation sets $T$ and $T \cup T^*$.}
    \label{fig:plot_6_5}
\end{figure}

\begin{figure}[htbp]
    \centering
    \resizebox{\columnwidth}{!}{
        \includegraphics{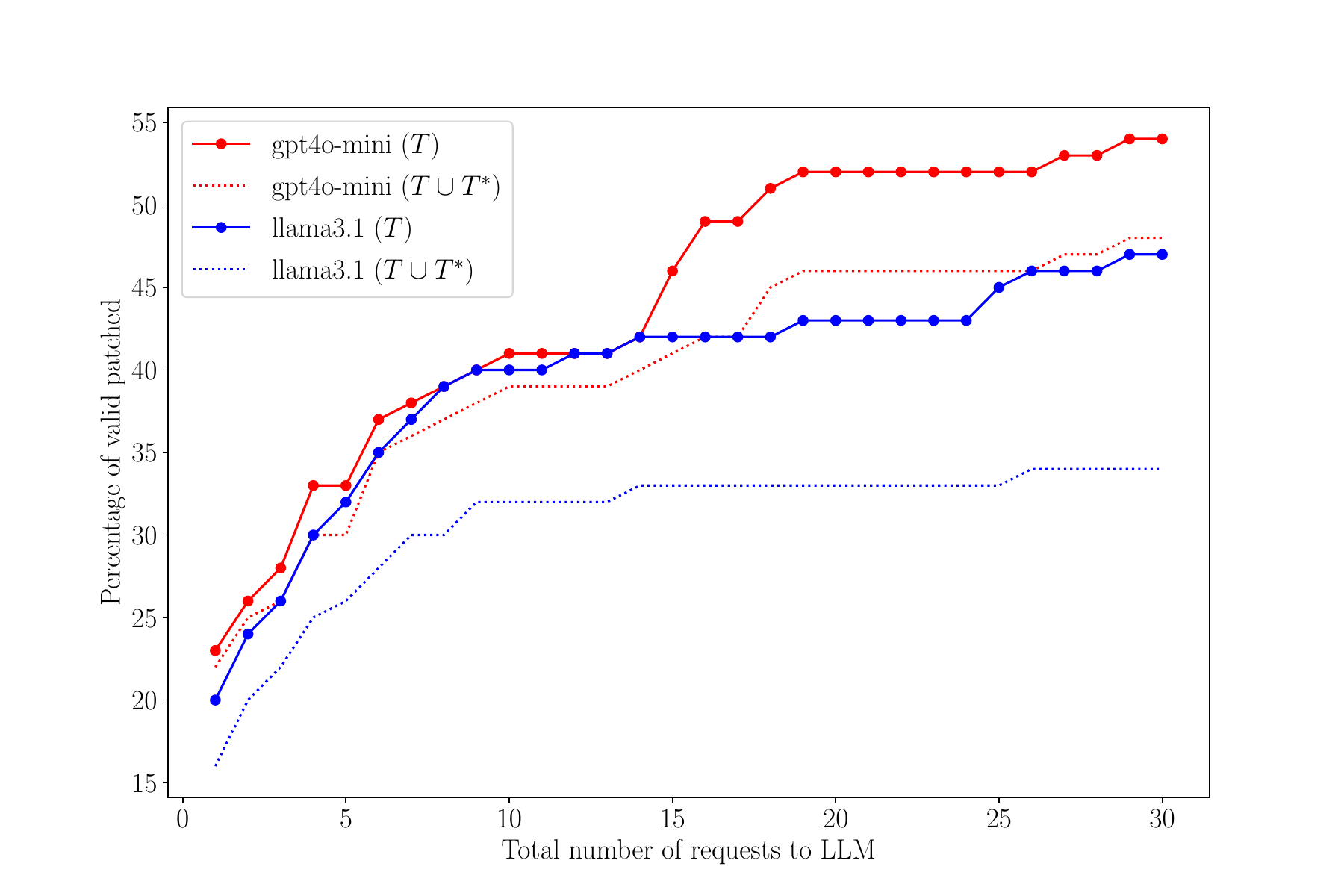}
    }
    \caption{Accumulated percent of valid patches generated during 30 repetitive independent patch generations. Two LLMs: llama3.1 70B Instruct and gpt4o-mini. Two patch validation sets $T$ and $T \cup T^*$.}
    \label{fig:plot_30_1}
\end{figure}

\section{Results \& Discussion}

\subsubsection*{RQ-1} The conversational experiment includes 6 consecutive conversation rounds; each round is limited by 5 requests to LLM. The  Figure \ref{fig:plot_6_5} plots the value of both metrics for both models. Metrics are calculated after each attempt.

The number of valid patches increases as total number of requests grows and reach 62\% for llama3.1 and 56\% for gpt4o-mini. The fastest grows takes place at the beginning of the experiment.

The number of valid patches that pass both public and hidden test suites drops to 47\% and 46\%  for llama3.1 and gpt4o-mini. The top 2 approaches for SWE-Bench Codestory Aide + Mixed Models and Honeycomb have solved 44\% among 92 problems participated in the experiment.

The \textit{simple} conversational patch generation pipeline allows to generate patches that are valid in 62\% cases; and in 47\% of cases the generated patches also pass hidden test suite.

The experiment version where both public $T$ and private $T^*$ suites are used simultaneously shown worse results. This result may be explained by the fact that private test suite size usually multiple times larger than public, and its failure output significantly increases the LLM request size.


\subsubsection*{RQ-2} The second experiment includes 30 consecutive rounds with a single request to LLM. It does not suppose of using any test failure feedback. The Figure \ref{fig:plot_30_1} plots experiment data.

The number of valid patches increases as total number of requests grows and reach 46\% for llama3.1 and 54\% for gpt4o-mini; but drops to 34\% and 47\% correspondingly to pass both test suites.

One interesting finding is that repetitive pipelines produced valid patches\footnote{for both public and private test suites} for six problems that were not addressed with conversational pipeline.

The experimental data shows that llama3.1 based patch generation with failure information (62\%) significantly outperforms repetitive design (46\%). Surprisingly, no differences were found for gpt4o-mini. A possible explanation for this might be that  gpt4o-mini is over-fitted on the experimental data and generates patches with equal effectiveness regardless of failure test information.

\section{Related Work}

\subsubsection*{Test Suite Based Program Repair}

Automatic Program Repair (APR) is a general task aims to help developers fix software defects. A test suite based program repair supposes to have at least one failing test. These tests can be helpful for all common repair steps (i) to localize the defect \cite{Yang2017Empirical, Le2012GenProg, Jiang2018Shaping} in the program; (ii) to generate a patch \cite{Ye2022SelfAPR}; (iii) and to validate the generated patch \cite{Le2012GenProg, Jiang2018Shaping}. This study uses failing tests for the last two steps.

\subsubsection*{Automatic Program Repair with pretrained LLMs}

Previous studies has shown that pre-trained LLMs are able to outperform state-of-the-art learning-based APR tools on both Java and Python \cite{Prenner2022CanCodex, Xia2023APR}.

The study \cite{xia2023conversationgoingfixing162} firstly introduces conversational-driven APR paradigm. A usage of LLM and unit test failure information outperforms learning-based and traditional APR tools. Similarly, the study \citep{zhang2024criticalreviewlargelanguage} demonstrates that conversational APR might be useful for a difficult-to-fix bugs. The current study evaluates potential of conversational patch generation on problems from SWE-Bench \cite{jimenez2024swebenchlanguagemodelsresolve}.

\subsubsection*{Automatic GitHub Issue Resolving}

Automatic GitHub Issue Resolving was introduced in \cite{jimenez2024swebenchlanguagemodelsresolve} to evaluate and challenge AI abilities. It is APR task at project level, with provided issue description and public and hidden test suites. This challenge got an attention and multiple approaches have been proposed \citep{yang2024sweagent, zhang2024autocoderover, tao2024magis}. The majority of them exploit pretrained LLM for fault localization and patch generation steps.

\subsubsection*{Repeated Sampling}
Today's state-of-the-art on SWE-Bench Lite is 43\%. The repetitive sampling \citep{brown2024largelanguagemonkeysscaling} allows to achieve up to 56\%. Authors sample with a high temperature 250 solutions for each SWE-Bench problem to enhance solution diversity and increase the chance of obtaining the correct one. In like manner, including test failure information in LLM request results in greater solution diversity.

\section{Conclusions}

The results of this study demonstrates that LLM-based conversational patch generation with a failing unit test feedback has potential to improve effectiveness of automatic program repair at project-level.

The naive conversational patch generation pipeline, being applied to problems from SWE-Bench Lite, allows to generate patches that pass 47\% of public and hidden tests. This is 7\% higher than current state-of-the-art. These findings opens future questions about the importance of precise fault localization, failing unit-tests, and more advanced patch generation pipelines.

\begin{Appendix}
\pagebreak
    \section{\\List of SWE-Bench problems in the experiment}
    sympy\_sympy-24066, sympy\_sympy-23191, matplotlib\_matplotlib-26011, mwaskom\_seaborn-3190, sympy\_sympy-22005, sympy\_sympy-18057, sympy\_sympy-13480, sympy\_sympy-12236, matplotlib\_matplotlib-24970, pydata\_xarray-4094, sympy\_sympy-21171, matplotlib\_matplotlib-23314, matplotlib\_matplotlib-25079, sympy\_sympy-14817, sympy\_sympy-22840, pylint-dev\_pylint-5859, sympy\_sympy-12481, sympy\_sympy-13177, scikit-learn\_scikit-learn-13584, matplotlib\_matplotlib-25433, sympy\_sympy-13895, sympy\_sympy-20049, sphinx-doc\_sphinx-8595, matplotlib\_matplotlib-23476, pydata\_xarray-3364, pylint-dev\_pylint-6506, scikit-learn\_scikit-learn-14894, astropy\_astropy-12907, pytest-dev\_pytest-11148, pylint-dev\_pylint-7993, matplotlib\_matplotlib-24149, scikit-learn\_scikit-learn-10949, sympy\_sympy-24213, sympy\_sympy-13437, pydata\_xarray-4248, pytest-dev\_pytest-11143, sympy\_sympy-13043, sympy\_sympy-13471, mwaskom\_seaborn-2848, sympy\_sympy-19254, astropy\_astropy-14995, sympy\_sympy-18621, sympy\_sympy-15609, sympy\_sympy-20154, sympy\_sympy-18532, sympy\_sympy-11897, sympy\_sympy-13647, sympy\_sympy-13146, sympy\_sympy-15346, sphinx-doc\_sphinx-8713, sphinx-doc\_sphinx-7975, sympy\_sympy-16503, sympy\_sympy-14317, sympy\_sympy-13971, sympy\_sympy-17630, sympy\_sympy-14308, sympy\_sympy-15678, matplotlib\_matplotlib-23299, scikit-learn\_scikit-learn-13779, pydata\_xarray-5131, sympy\_sympy-24152, scikit-learn\_scikit-learn-12471, sympy\_sympy-18835, matplotlib\_matplotlib-23563, mwaskom\_seaborn-3010, sphinx-doc\_sphinx-7738, astropy\_astropy-6938, sympy\_sympy-21847, sympy\_sympy-21612, sympy\_sympy-18189, sympy\_sympy-13915, sympy\_sympy-14774, astropy\_astropy-7746, mwaskom\_seaborn-3407, matplotlib\_matplotlib-23987, sympy\_sympy-21627, sympy\_sympy-22714, sympy\_sympy-20212, sympy\_sympy-16281, sympy\_sympy-17139, sphinx-doc\_sphinx-8721, sympy\_sympy-16988, pydata\_xarray-4493, matplotlib\_matplotlib-24334, scikit-learn\_scikit-learn-15535, sympy\_sympy-24102, pytest-dev\_pytest-7168, sympy\_sympy-21379, scikit-learn\_scikit-learn-14087, sympy\_sympy-23262, scikit-learn\_scikit-learn-13142, sympy\_sympy-16792

\begin{tikzpicture}
    \begin{axis}[
        ybar,
        width=12cm,
        height=8cm,
        bar width=0.5cm,
        ylabel={SWE-Bench problems by project},
        symbolic x coords={sympy, matplotlib, mwaskom, pydata, pylint-dev, scikit-learn, astropy, pytest-dev, sphinx-doc},
        xtick=data,
        x tick label style={rotate=45, anchor=east},
        nodes near coords,
        ymin=0,
        enlarge x limits={abs=1cm},
    ]
    \addplot coordinates {(sympy, 52) (matplotlib, 13) (mwaskom, 4) (pydata, 4) (pylint-dev, 4) (scikit-learn, 8) (astropy, 4) (pytest-dev, 3) (sphinx-doc, 4)};
    \end{axis}
\end{tikzpicture}
\end{Appendix}
\begin{Appendix}
    \section{\\Prompt Templates}
    \subsection{\\Prompt template A}
    \begin{tcolorbox}[colframe=black, width=\textwidth, arc=0mm, auto outer arc]
        You need to fix the issue:\\
        \textbf{<ISSUE\_DESCRIPTION>}
            \\You are professional Python programmer. You are asked to fix code. Write the answer ONLY in the following format:\\
        <replace>\\
         \{fixed python code\}\\
        </replace>,\\ 
        where the code after <replace> should be the code which was already rewritten.
        \\Fix the following code, insert fixed code of the full function, do not shorten it.\\
        \textbf{<FUNCTION\_TO\_FIX>}
    \end{tcolorbox}

    \subsection{\\Prompt template B}
    \begin{tcolorbox}[colframe=black, width=\textwidth, arc=0mm, auto outer arc]
        Your code has errors, make reasoning then fix function using <replace></replace> tags. Replace function which was sent to you. You need to change only provided function, you must not add new functions or new classes. \\ 
        \textbf{<FAILURE LOG>}
    \end{tcolorbox}
    \subsection{\\Prompt C}
    \begin{tcolorbox}[colframe=black, width=\textwidth, arc=0mm, auto outer arc]

        Error: Failed to create or apply patch. Use <replace></replace> tags. Rewrite what you fixed in the last iteration using the required format. Replace function which was sent to you. You need to change only provided function, you must not add new functions or new classes!
    \end{tcolorbox}
    
\end{Appendix}

\bibliographystyle{plainnat}
\bibliography{references}

\end{document}